# Photoacoustic vector tomography for deep hemodynamic imaging

Yang Zhang[1,†], Joshua Olick-Gibson [1,†], Anjul Khadria[1], Lihong V. Wang[1,2,*]

[1] Caltech Optical Imaging Laboratory, Andrew and Peggy Cherng Department of Medical Engineering, California Institute of Technology, 1200 East California Boulevard, Pasadena, CA 91125, USA.

[2] Caltech Optical Imaging Laboratory, Department of Electrical Engineering, California Institute of Technology, 1200 East California Boulevard, Pasadena, CA 91125, USA.

[†]These authors contributed equally to this work: Yang Zhang, Joshua Olick-Gibson

*Corresponding author. L. V. Wang (LVW@caltech.edu)

## Abstract

Non-invasive imaging of deep blood vessels for mapping hemodynamics remains an open quest in biomedical optical imaging. Although pure optical imaging techniques offer rich optical contrast of blood and have been reported to measure blood flow, they are generally limited to surface imaging within the optical diffusion limit of about one millimeter. Herein, we present photoacoustic vector tomography (PAVT), breaking through the optical diffusion limit to image deep blood flow with speed and direction quantification. PAVT synergizes the spatial heterogeneity of blood and the photoacoustic contrast; it compiles successive single-shot, wide-field photoacoustic images to directly visualize the frame-to-frame propagation of the blood with pixel-wise flow velocity estimation. We demonstrated *in vivo* that PAVT allows hemodynamic quantification of deep blood vessels at five times the optical diffusion limit (more than five millimeters), leading to vector mapping of blood flow in humans. By offering the capability for deep hemodynamic imaging with optical contrast, PAVT may become a powerful tool for monitoring and diagnosing vascular diseases and mapping circulatory system function.

## Introduction

*In vivo* hemodynamic imaging provides invaluable functional information in tissues and organs. Biomedical optical imaging techniques, such as Doppler-based methods[1] (laser Doppler flowmetry[2], Doppler optical coherence tomography[3]), speckle-based methods[4] (laser speckle contrast imaging[5]), and red blood cell tracking methods (confocal microscopy[6]), have been reported to map the dynamics of blood flow due to their intrinsic optical contrast of blood. However, these techniques suffer from shallow depth penetration due to strong ballistic light attenuation in biological tissue. Thus, imaging blood flow with fine spatial resolution at depths beyond the optical diffusion limit (>1 mm)[7] by pure optical imaging remains challenging. Unlike pure optical imaging modalities, photoacoustic tomography (PAT) combines optical absorption contrast with low scattering ultrasonic detection to enable high spatiotemporal resolution and deep tissue imaging of endogenous chromophores, such as hemoglobin[8]. As such, PAT is uniquely positioned as a functional, anatomical, and molecular imaging modality which can image blood vessels beyond the optical diffusion limit with fine spatial resolution[7].

PAT primarily has three main forms of implementation: optical resolution photoacoustic microscopy (OR-PAM), acoustic resolution photoacoustic microscopy (AR-PAM), and photoacoustic computed tomography (PACT)[7]. While all three techniques have excellent optical contrast for imaging blood vessels, thus far, only OR-PAM has been demonstrated to measure *in vivo* blood flow[9]. OR-PAM uses point-by-point scanning of an optical focus to enable transverse spatial resolution on the order of microns[7]. As such, OR-PAM can measure blood flow by tracking individual red blood cells (RBCs), whose diameters range from 7-8 microns[10]. However, this measurement is limited to the optical diffusion limit in biological tissue due to the inability to focus light beyond one transport mean free path[11]. AR-PAM employs raster scanning of an acoustic focus for three-dimensional (3D) imaging at acoustic resolution[11]. Since the resolution in AR-PAM is sacrificed to image deeper than OR-PAM, deep blood flow measurement is challenging because individual RBCs cannot be resolved. Similar in principle to AR-PAM, acoustic resolution photoacoustic doppler flowmetry (AR-PAF) uses wide-field illumination and a single-element focused transducer to calculate A-line cross-correlations between photoacoustic (PA) waveforms. Previous work has demonstrated that *ex vivo* whole blood velocity measurement is possible for high center frequency ($f_c$) transducers ($f_c \geq 30$ MHz) using AR-PAF[12], but *in vivo* blood flow measurement beyond the optical diffusion limit remains an outstanding hurdle. Conversely, PACT uses wide-field illumination coupled with an array of ultrasonic detectors to image vasculature at depths beyond the optical diffusion limit with acoustic resolution[13]. Generally speaking, measuring blood flow in deep tissue with PACT has been challenging because 1) compared to OR-PAM,

PACT has a lower resolution, preventing it from being able to resolve individual RBCs; 2) the photoacoustic signals within the lumen of a vessel are suppressed relative to the signals at the boundaries due to the random summation of absorption signals from millions of RBCs in each lumen imaging voxel, thus rendering PACT as "speckle-free"[14]. While recent work has utilized PACT speckle field decorrelation to extract velocity measurements in ink phantoms and chicken embryos, direct imaging of blood flow beyond the optical diffusion limit has remained elusive[15].

In this article, we present photoacoustic vector tomography (PAVT) as a framework to achieve, to our knowledge, the first vector maps of human blood flow by photoacoustics beyond the optical diffusion limit. The key features of this framework can be summarized as follows: 1) The synergy between the spatial heterogeneity of blood and the photoacoustic contrast produces strong photoacoustic signals in the lumen of the blood vessels. 2) Successive compilation of single-shot, wide-field photoacoustic images allows direct visualization of the blood flow throughout the lumen. 3) Applying pixel-wise motion estimation algorithms to the reconstructed images generates blood flow vector maps with speed and direction quantification. Through simulation and phantom validation, we demonstrated that PAVT blood flow measurement is facilitated by the heterogeneity of the blood. The accuracy of this measurement was assessed through trials on *ex vivo* blood in which measured flow speeds were validated against known values. We acquired images in the vessels of the hand and arm regions of healthy subjects using a linear ultrasound array probe coupled to an optical fiber bundle. We demonstrated *in vivo* that vector flow maps could be obtained in blood vessels greater than 5 mm in depth. Moreover, we demonstrated the potential for vector flow imaging to measure unique flow patterns at irregular interfaces in the blood vessel, such as valve regions. Lastly, we measured the *in vivo* functional PA blood flow responses to the inflation and release of a blood pressure cuff. This work establishes PAVT as a viable imaging technique for monitoring and diagnosing vascular diseases and mapping circulatory system function.

## Results

The principle of photoacoustic vector tomography (PAVT) for deep hemodynamic imaging is illustrated in Fig. 1. Fig. 1a shows the schematic of the system. An ultrasonic probe is placed on top of the skin, and a laser delivers light through a fiber bundle to the blood vessels. The hemoglobin content within RBCs absorbs the light and converts it into heat. Based on the photoacoustic effect, acoustic waves are generated through thermoelastic expansion and detected by the ultrasonic probe. The data are then streamed to the computer through the data acquisition (DAQ) module, and we employ the universal back projection algorithm[16] to reconstruct the images. Fig. 1b illustrates the effect of blood heterogeneity (RBC spatial distribution) on the visibility of the signals arising from the lumen region of the blood vessel. For uniform blood, the simulated photoacoustic signals from the lumen region are suppressed relative to the signals at the boundaries (known as the boundary buildup effect[14]), resulting in little visibility in the lumen region of the image, whereas in nonuniform blood, the simulated photoacoustic signals from the lumen region show an increased amplitude, resulting in heightened visibility in the lumen region relative to the uniform case (simulation details can be found in the Methods). Fig. 1c is a representative *in vivo* photoacoustic image of a blood vessel located beyond 1 mm depth. Similar to the simulated image in the nonuniform case, the *in vivo* image shows detectable signals from the lumen region of the blood vessel. After acquiring multiple frames of data, we can clearly visualize the propagation of the lumen signals from frame-to-frame, as shown in the magnified view of the region in the white box in Fig. 1c (further data processing details can be found in the Methods). We then apply a pixel-wise flow estimation algorithm to the images, extracting both the direction and magnitude of the blood flow inside the vessel. As shown in Fig. 1d, this map shows a laminar flow pattern in the lumen region that has a higher speed through the center of the vessel and a lower speed at the edges.

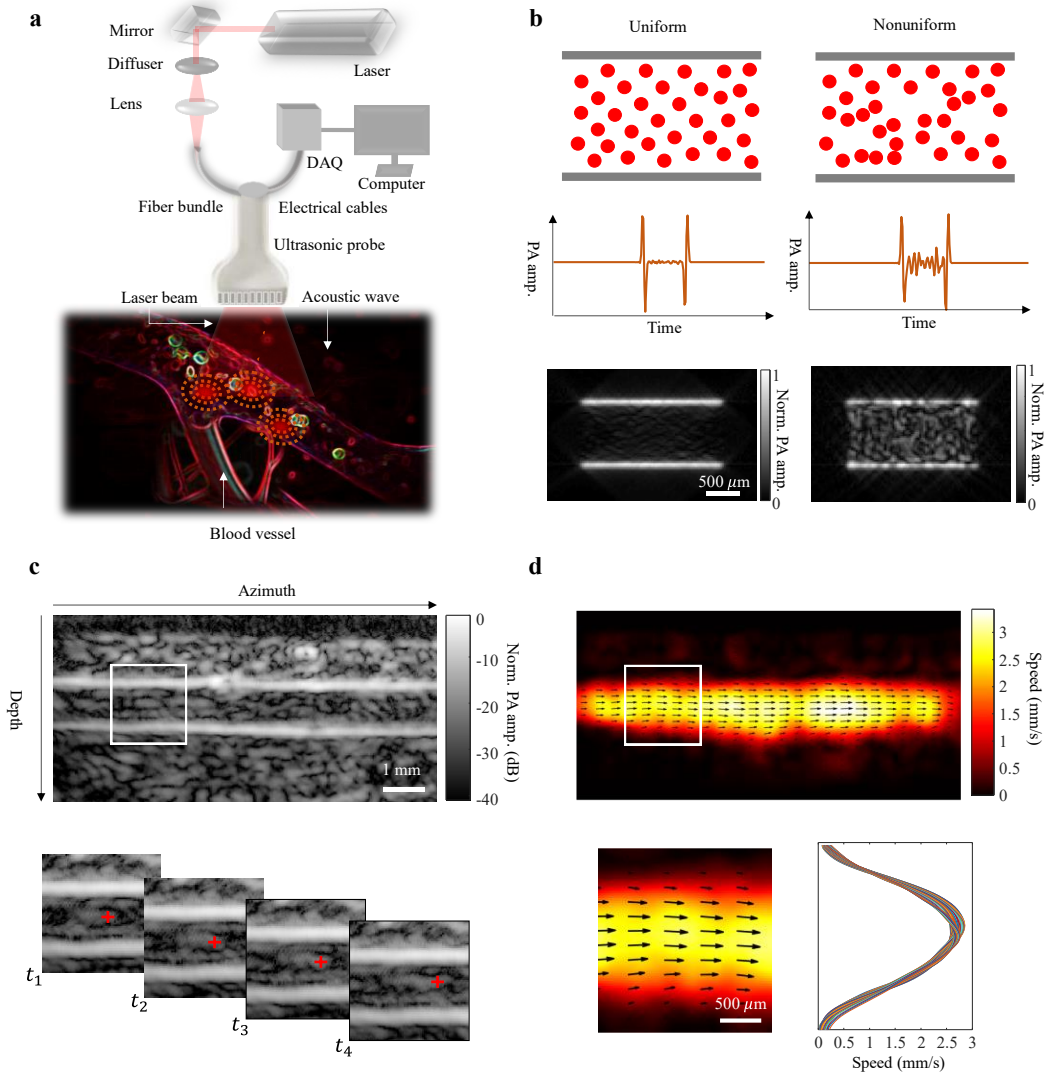

**Fig. 1 | Photoacoustic vector tomography (PAVT). a**, Schematic of the imaging system. Light from a laser is delivered to the blood vessel region via a fiber bundle. Photoacoustic waves are generated and then detected by the ultrasonic probe. Photoacoustic signals are streamed to the computer through the data acquisition (DAQ) module. **b,** Effect of blood heterogeneity on the visibility of the signals arising from the lumen region of the blood vessel. For uniform blood, the simulated photoacoustic signals from the lumen region are suppressed relative to the signals at the boundaries, resulting in little visibility in the lumen region of the image, whereas in nonuniform blood, the simulated photoacoustic signals from the lumen region show an increased amplitude, resulting in heightened visibility in the lumen region relative to the uniform case. **c**, Structural images of a blood vessel are reconstructed in real-time. Magnified view of the region in the white box in **c** (top row) shows that the image features in the lumen region (beyond 1 mm depth) can be tracked in successive images at time points $t_1$ to $t_4$. **d**, After tracking all of the features in the field of view, a blood flow speed map is reconstructed. Quantitative representation of the flow vector fields is overlayed on the speed map. Magnified view of the region in the white box in **d** (top row) shows that the flow has a laminar pattern.

To characterize the system resolution, we imaged a lead point source (actual diameter 50 µm). Fig. 2a shows an axial resolution of 125 µm and a lateral resolution of 150 µm. To determine the mechanism for flow detectability and validate the speed estimation accuracy of PAVT, we designed a blood flow phantom to closely mimic *in vivo* circulation. Fig. 2b shows the set-up for the flow phantom, in which a converging bifurcation is fabricated by two inlet channels ($C_1$ and $C_2$) feeding into an outlet channel ($C_3$). While imaging the outlet channel, five measured speeds were validated against preset syringe flow speeds. The results in Fig. 2c show strong agreement between the measured and preset values for all five speeds. The mechanism for flow detectability is demonstrated in Fig. 2 d-i. Fig. 2 d,f,h show the structural images of channels $C_1$, $C_2$, and $C_3$, respectively, before (top) and after (bottom) the induction of pressure fluctuations in the inlet channels, whereas Fig. 2 e,g,i show the corresponding vector flow maps for each channel and scenario. Fig. 2 d-g show suppressed lumen signals and no flow detectability (97% relative error for measured mean speeds of 0.1 mm/s versus ground truth of 3.3 mm/s) in the inlet channels $C_1$ and $C_2$ with and without pressure fluctuations. In the top panel of Fig. 2h, the structural image for outlet channel $C_3$ shows a detectable signal in the center of the channel, which likely arises from the mixing of blood from each inlet channel, thus producing a detectable nonuniformity at the center of the outlet channel. The vector flow image in the top panel of Fig. 2i shows weak flow detectability (80% relative error for measured mean speed of 0.6 mm/s versus ground truth of 3.3 mm/s). After the induction of pressure fluctuations in the inlet channels, the structural image in the bottom panel of Fig. 2h shows the nonuniformity dispersed throughout the lumen region allowing us to clearly track the flow (2.5% relative error for measured mean speed of 3.4 mm/s versus ground truth of 3.3 mm/s) in the vector flow map shown in the bottom panel of Fig. 2i. Thus, we conclude that the visibility of the photoacoustic lumen signals is enhanced by spatial heterogeneity in the blood.

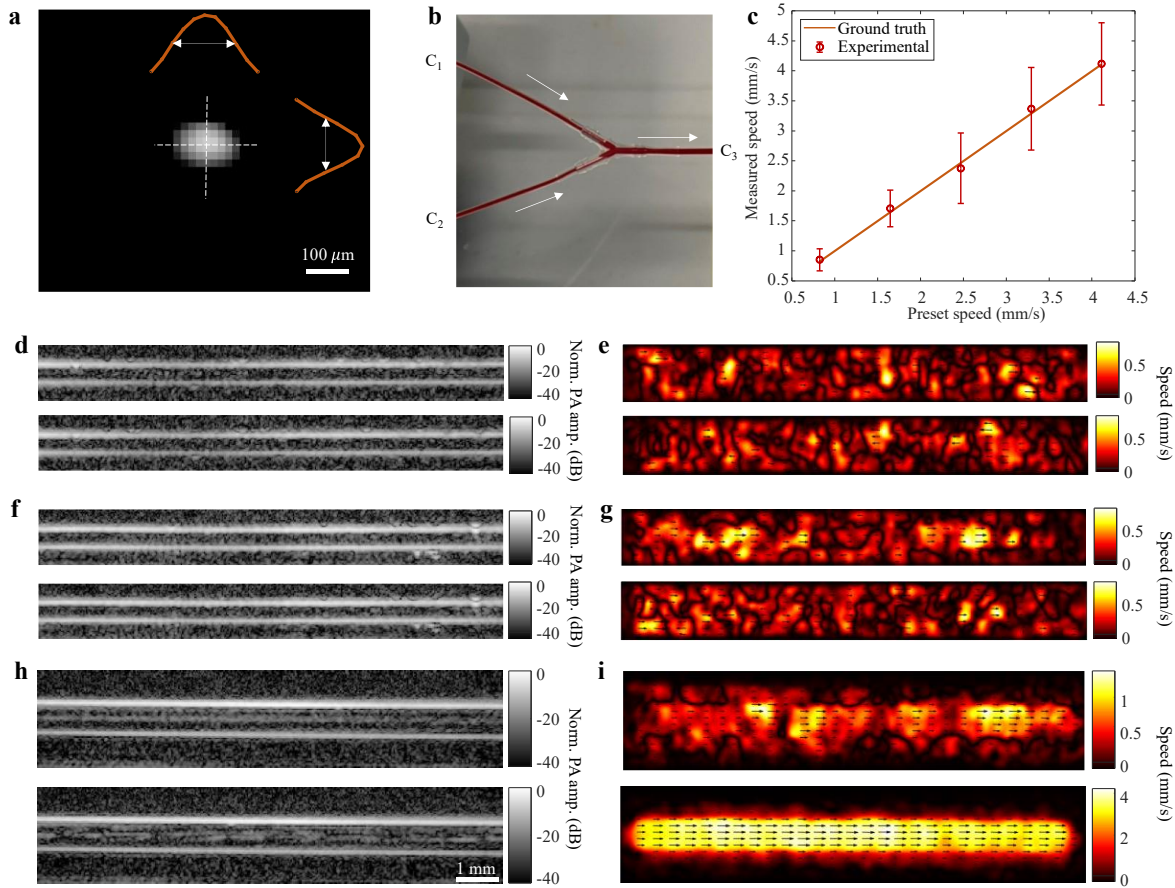

**Fig. 2 System characterization and phantom validation. a,** Image of a lead point source (actual diameter 50 µm), showing axial and lateral resolutions of 125 and 150 µm, respectively. **b,** Blood phantom set-up with inlet channels $C_1$, $C_2$, and outlet channel $C_3$. **c,** Measured flow speeds plotted against preset syringe flow speeds. The mean ± standard deviation (in mm/s) speeds were (from left to right): 0.85 ± 0.18, 1.71 ± 0.31, 2.38 ± 0.59, 3.37 ± 0.69, and 4.12 ± 0.69 with relative errors of 3.7, 3.9, 3.6, 2.5, and 0.2 %, respectively ($n$ = 6232). Error bars represent the s.d. for each measurement. **d,f,h,** Structural images in channels $C_1$, $C_2$, and $C_3$, respectively, before (top) and after (bottom) the induction of pressure fluctuations. **e,g,i,** Vector flow maps in channels $C_1$, $C_2$, and $C_3$, respectively, before (top) and after (bottom) the induction of pressure fluctuations.

Representative PAVT images of four blood vessels at depths beyond the optical diffusion limit are shown in Fig. 3. Fig. 3a shows a vessel in the wrist imaged at an optical wavelength (λ) of 1,064 nm and a pulse repetition frequency (PRF) of 100 Hz. The vessel has a varying diameter along its longitudinal axis with a high flow speed in the upstream narrower region (likely due to smooth muscle contraction) and a lower flow speed in the downstream dilated region. Fig. 3b shows a vessel in the palmar region imaged at a PRF of 100 Hz with an isosbestic wavelength (805 nm) of oxy- and deoxyhemoglobin. Demonstration of flow measurements at other wavelengths can be found in Supplementary Fig. 1. In Fig. 3c-d, the structural images acquired at λ = 1,064 nm and

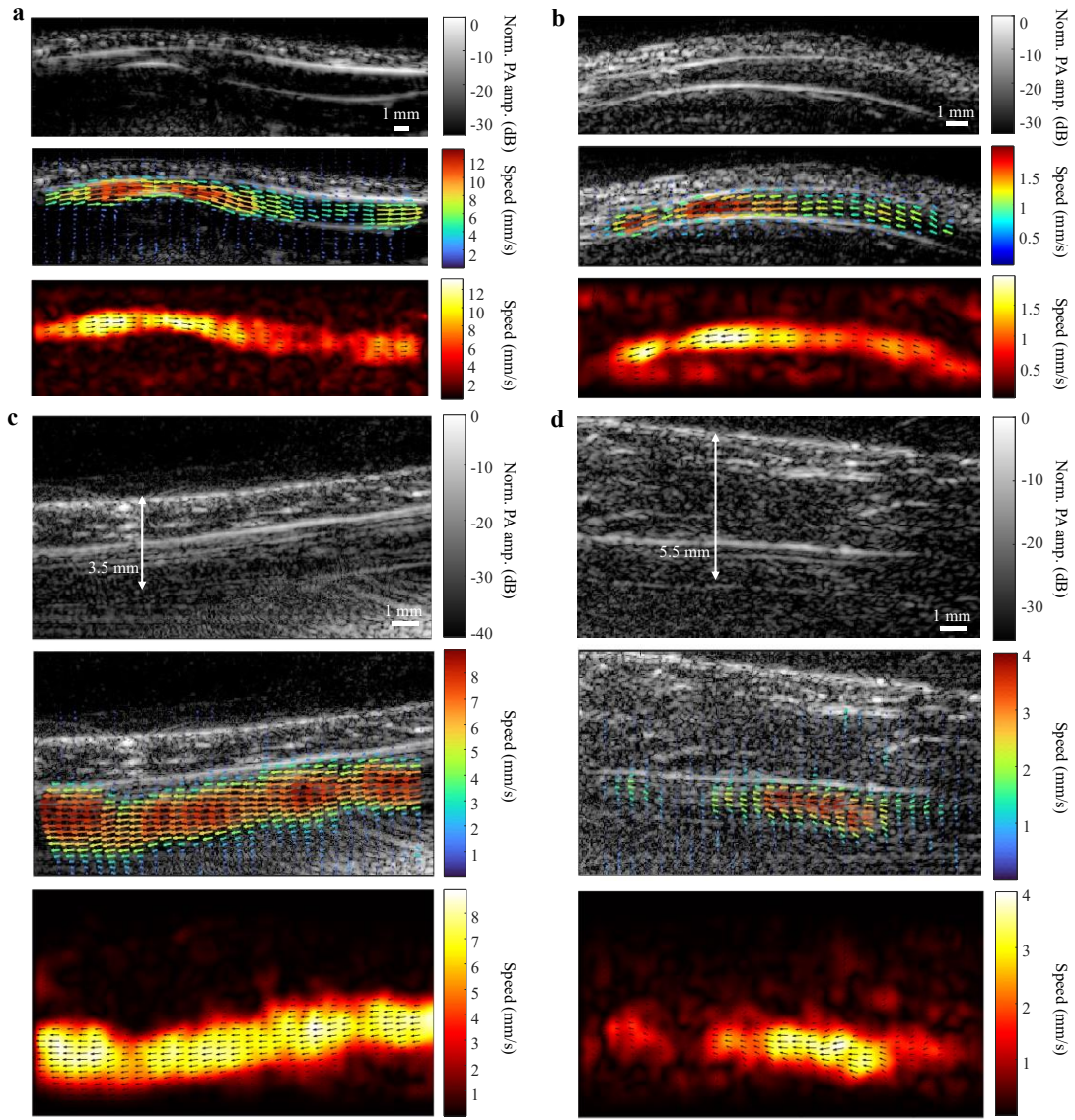

**Fig. 3 | PAVT vector flow maps. a**, Structural image (top), overlayed vector flow map (middle), and vector flow map (bottom) of a vessel in the wrist showing dilation in the rightmost region. **b**, Structural and vector maps of a vessel in the palmar region imaged at an isosbestic wavelength for oxygenated and deoxygenated hemoglobin (805 nm). **c-d**, Structural and vector maps of vessels in the forearm imaged at 3.5 and 5.5 mm depths, respectively.

PRF = 20 Hz show that the distances from the skin surfaces to the axes of the vessels are 3.5 mm and 5.5 mm, respectively, with the vector maps confirming that the blood flow is still detectable. Furthermore, we demonstrated that PAVT could consistently measure blood flow through varying depths of light-scattering tissue (see Supplementary Fig. 2).

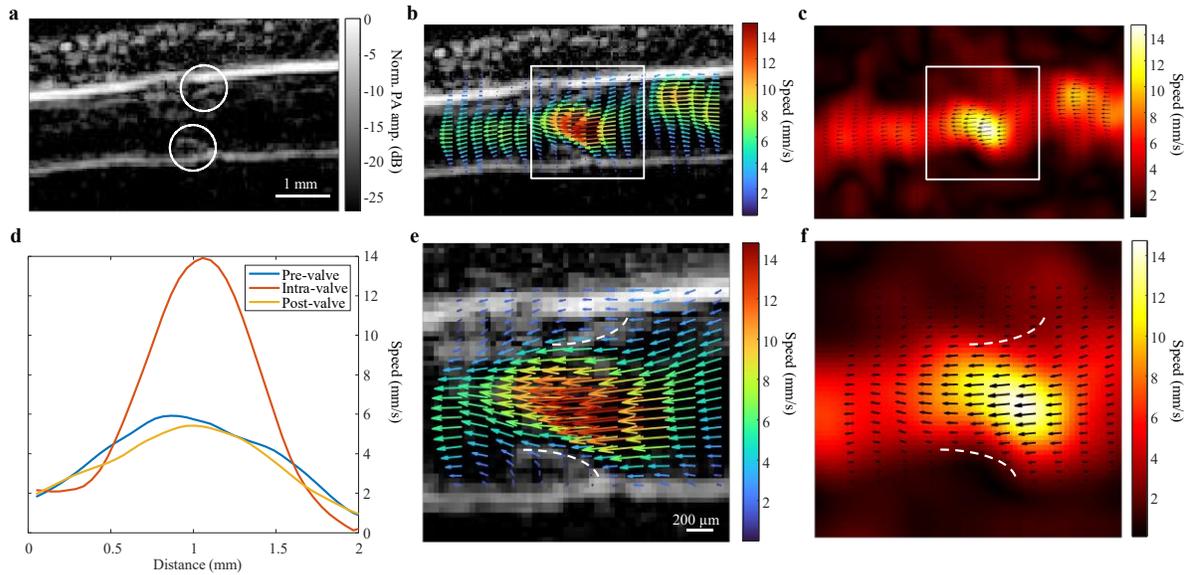

**Fig. 4 | PAVT characterization of hemodynamics within a valve. a,** A venous valve in the carpal tunnel region in the arm (valve boundaries indicated by white circles). **b-c,** Vector maps showing a higher speed inside the valve than the vessel regions upstream and downstream to the valve. **d,** Speed profiles taken upstream, within, and downstream the valve (pre-valve, intra-valve, and post-valve, respectively). **e-f,** Magnified views of the maps in **b-c** with the valve boundaries indicated by the dashed white lines.

A carpal tunnel vein with a valve is shown in Fig. 4, as indicated by the white circles in Fig. 4a. The vector maps and speed profiles in Fig. 4b-c and Fig. 4d, respectively, indicate a higher speed in the narrow diameter intra-valve region and lower speeds in the wider diameter regions surrounding the valve. In Fig. 4e-f, magnified views of the regions indicated by the white boxes show a converging flow pattern within the valve. Furthermore, the magnified vector maps highlight the sinus pocket regions[17] above and below the valve boundaries, as shown by the dashed lines in Fig. 4e-f.

To evaluate functional changes, we measured a subject's blood flow before, during, and after inflating a blood pressure cuff. The cuff was placed against the brachial vein in the upper left arm, and the imaged vessel was located in the metacarpal region, distal to the application site of the cuff. Fractional speed changes were measured relative to the baseline flow. Fig. 5 shows a flow speed decrease of approximately 70% while the cuff was inflated, followed by a transient flow speed increase of approximately 350% upon release of the cuff and eventually a steady-state return to the baseline. The 70% drop can be explained by incomplete cuffing, whereas the transient increase could be due to an immediate pressure release in the metacarpal vessel.

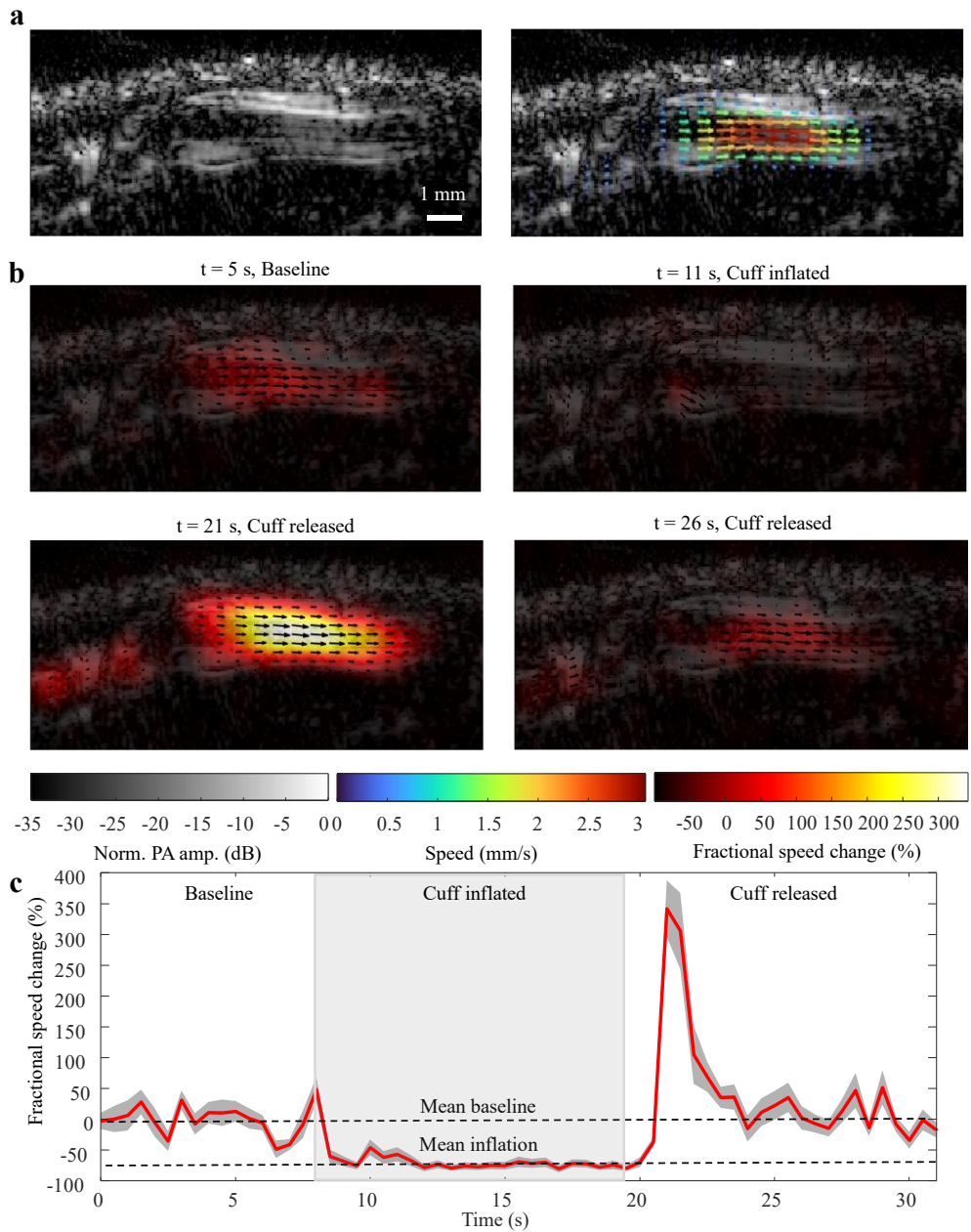

**Fig. 5 | Measuring functional responses to a blood pressure cuff.** A blood pressure cuff was applied to the upper left arm across the subject's brachial vein. Baseline flow was measured in a distal metacarpal vein, after which subsequent flow measurements were quantified relative to the mean of this baseline. The cuff was inflated for approximately 10 seconds before release. **a,** Structural image of the vessel (left) and the vector overlay map for the baseline flow speed (right). **b**, Vector maps for relative flow measurements at four different time points throughout the experiment: the baseline flow (top left), the flow during cuff inflation (top right), the flow during release (bottom left), and the flow after release (bottom right). **c**, Discrete-time sequence for relative flow speed changes. The flow speed decreased by approximately 70% during the application of the cuff and increased to approximately 350% upon release before returning to baseline.

## Discussion

This study introduces PAVT, which demonstrates the capability of photoacoustics to image human blood flow beyond the optical diffusion limit. Through phantom experiments, we validate the mechanism and accuracy of this measurement. By employing *in vivo* vector flow analysis, we demonstrate the versatility of PAVT across laser pulse repetition frequency (PRF), laser wavelength, and depths. Furthermore, we show that in the vicinity of valves, PAVT can measure flow patterns that deviate from typical laminar flow. In tracking the flow speed changes in response to a blood pressure cuff, this work establishes PAVT as a powerful imaging technique that can perform blood flow functional measurements in deep vessels. As such, PAVT outperforms existing pure optical methods for deep hemodynamic imaging and complements ultrasound imaging[18–20] by simultaneously providing rich molecular contrast[21].

There is an inherent tradeoff between imaging depth and laser PRF. For instance, the American National Standards Institute's (ANSI) safety limit for skin exposure at a laser wavelength of 1,064 nm is 50 mJ/cm$^2$ and 10 mJ/cm$^2$ at a laser PRF of 20 Hz and 100 Hz, respectively[22]. Therefore, if one wishes to maximize laser fluence in order to image deep blood vessels with a high signal-to-noise ratio (SNR), one must necessarily decrease the laser PRF to satisfy the ANSI limit. This tradeoff is exemplified in Fig. 3, in which we used a 20 Hz laser PRF to image blood flow at a depth of approximately 5.5 mm (Fig. 3d), whereas we used a 100 Hz laser PRF to image blood flow at a depth of approximately 1.5 mm (Fig. 3a). The effect of the frame rate on speed estimation is further expounded upon in Supplementary Fig. 3. Starting with a dataset acquired at a 100 Hz PRF over 1 second, we downsampled the images at effective frame rates of 50, 20, 10 and 5 Hz. The mean speeds showed reasonable agreement (within one standard deviation of the 100 Hz upsampled mean speed), at frame rates of 50, 20, and 10 Hz, with an underestimation at 5 Hz (relative error of 89%). These results are consistent with Equation 8, in which the calculated flow pattern correlation length and flow speed of approximately 0.6 mm and 2 mm/s, respectively, correspond to a minimum required laser PRF of 6.7 Hz according to the Nyquist sampling theorem[23].

To clearly visualize the flow, we employed two post-processing techniques prior to velocity quantification. First, we implemented a singular value decomposition (SVD)-based spatiotemporal filter to the reconstructed photoacoustic images to extract the dynamic features (i.e., blood flow). Next, we applied amplitude-based logarithmic compression[24] to the filtered images to highlight the lumen signals. As shown by the dashed lines in Supplementary Fig. 7, each of these steps may be implemented individually or together. The accuracy of these methods is demonstrated in Supplementary Fig. 4, in which we consider a representative phantom dataset. The relative errors for Supplementary Fig. 4a-d were 29, 6.9, 1.4, and 0.3%, respectively, indicating that implementing SVD and logarithmic compression yields the most accurate results, albeit with modest improvement over implementing logarithmic compression without SVD.

Additionally, we applied a pixel-wise noise floor filtering method (see details in Methods) before generating the final vector flow maps. The effect of this procedure is demonstrated in Supplementary Fig. 5, in which we consider a phantom dataset with a mean flow speed of 4.11 mm/s as the ground truth. Supplementary Fig. 5a shows the estimated speed map when averaging across all frames for each pixel, yielding a calculated mean speed of 3.93 mm/s (relative error of 4.5%). This underestimation is because when calculating pixel-wise frame-to-frame motion, low-speed noise measurements may erroneously be incorporated into the velocity quantification due to spatial gaps in the propagating flow pattern (as shown in the bottom panel of Fig. 2h). To avoid this underestimation, we calculated the temporal standard deviation of each pixel's speed measurements to form a pixel-wise noise floor, below which frames with low speeds (i.e., measurements due to noise) were excluded. Supplementary Fig. 5b shows the estimated speed map when averaging across the remaining frames, yielding a calculated mean speed of 4.07 mm/s (relative error of 0.9%).

An important aspect of PAVT is that it can detect variations in flow patterns throughout the lumen of the blood vessel. Fig. 4 highlights key features regarding the detectability of PAVT in blood vessel regions containing valves. From a structural standpoint, it is important to note that the venous extremities valves, such as the one imaged in this work, are composed of endothelial cells and connective tissue[25], for which photoacoustic imaging does not have high contrast relative to hemoglobin. However, like the walls of the blood vessels, the valve can be considered as a sub-boundary of the blood, and we can detect the sub-boundaries of these structures due to the boundary buildup effect. Our PAVT images reveal that we can detect complex vector flow patterns in the regions surrounding the valves. Furthermore, our speed profiles in Fig. 4d are consistent with that predicted by the continuity equation[26], showing a higher speed in the narrow diameter intra-valve region and lower speeds in the wider diameter regions before and after the valve (relative error of 3.2%).

Here, we consider the most likely hypothesis for the mechanism of blood flow detectability by PAVT. We designed a flow phantom (see Fig. 2) to closely mimic *in vivo* circulation, consisting of a converging bifurcation that has two inlet channels and an outlet channel with diameters roughly corresponding to Murray's law[27] (i.e., the cubic of the vessel radius $r$ is conserved: $r_{\text{out}}^3 = r_{\text{in1}}^3 + r_{\text{in2}}^3$, where "out" denotes the outgoing branch and "in1" and "in2" denote the two incoming branches, respectively). Because whole blood is uniform, we were unable to image baseline blood flow in either of the inlet channels. However, as shown in the top panel of Fig. 2h, the converging bifurcation produced a detectable nonuniformity of RBC distribution in the center of the outlet channel. Furthermore, inducing pressure fluctuations in the inlet channels resulted in the dispersion of this nonuniformity throughout the lumen of the outlet channel, allowing us to clearly visualize the blood flow. From a physiological perspective, the bifurcation structure naturally occurs throughout the circulatory system, while the pressure fluctuations can be induced by the opening and closing of one-way valves in the veins, as well as from smooth muscle contractions. In validating the flow speed measurements with the preset syringe flow speeds, we hypothesize that the two aforementioned factors facilitate *in vivo* flow measurement by inducing blood spatial heterogeneity that is detectable at our probe's center frequency.

For completeness, we also consider the hypothesis that PAVT flow measurement is facilitated by the propagation of oxygen saturation (sO$_2$) heterogeneities. This hypothesis is reasonable due to PA's selective contrast to oxy- and deoxyhemoglobin at most wavelengths. However, as shown in Fig. 3, we were able to measure flow even at an isosbestic wavelength of 805 nm, effectively ruling out this hypothesis. Moreover, we demonstrated that PAVT blood flow measurement is versatile, with consistent speed measurements across three different laser wavelengths (750 nm, 805 nm, and 900 nm) in the near-infrared (NIR) spectral region (see Supplementary Fig. 1) while imaging a palmar vein. Using linear unmixing with surface fluence compensation[28], we calculated an sO$_2$ of ~70%, which agrees with accepted literature values in healthy subjects[29], indicating that spectral PAVT measurement of blood flow and sO$_2$ is achievable. Simultaneous measurement of blood flow and sO$_2$ can provide crucial physiological information regarding brain function[30] and hallmarks of cancer, such as angiogenesis and hypermetabolism[31].

One limitation of this study is that we only imaged blood flow in veins. Supplementary Fig. 6 shows two representative arterial images. As demonstrated by the speed maps, we could not visualize the arterial blood flow and generate a reasonable vector flow map for either vessel. Based on our simulation and phantom experiments, we believe that the reason is that the blood in the arteries is more uniform than that in the veins. From a physiological perspective, this makes sense because by the time blood from the venous circulation returns to the heart and transitions to the arterial circulation, the blood has become well-mixed. As such, the spatial heterogeneity that highlights the photoacoustic signals in the venous blood is no longer present. Future work should seek to overcome this physiological hurdle and explore the feasibility of PAVT for arterial blood flow measurement.

Future work should also focus on extending PAVT to other regions of the body, as well as using this capability to diagnose and monitor hemodynamic pathologies. In particular, PAVT may be used to measure functional responses in the brain[32]. PAVT may also be used to diagnose chronic venous insufficiency[33], in which improper functioning of valves in the legs causes blood to pool in the lower extremities, which can result in severe circulatory complications due to a decrease in the venous blood supply to the heart. Furthermore, in principle, PAVT is not limited to the linear array probe that we employed in this work. As such, future work may extend this technique to ring array and 3D ultrasonic array geometries that allow for a more complete characterization of organ systems' hemodynamics.

**Methods**

**System construction.** We employed a 256-element linear ultrasonic transducer array (LZ250, VisualSonics Inc.; 13-24 MHz bandwidth) for photoacoustic signal detection. The transducer array has a size of 23 mm × 3 mm, and each element of the array has a cylindrical focus with a 15 mm focal length. The ultrasound probe was directly connected to the 256 DAQ channels of the Verasonics Vantage 256 system (Verasonics Inc.; 14-bit A/D converters; 62.5 MHz sampling rate; programmable gain up to 51 dB) through a UTA 360 connector. The photoacoustic signals were acquired and digitized into local memory and then transferred to a host computer via PCI express.

For light delivery, we incorporated an optical fiber bundle with the ultrasound probe. The fiber bundle and the array were coaxially aligned to maximize the system's performance. In the front end of the fiber bundle, a 1,064 nm wavelength laser beam (Quantel Brilliant B pulsed YAG laser;

10 Hz; 5- 6 ns pulse width; Quantel Q-Smart 450 laser; 20 Hz; 5-6 ns pulse width) or a 670 nm – 1070 nm laser beam (SpitLight EVO III, InnoLas Laser Inc.; 100 Hz; 5-8 ns pulse width) was utilized to pass through the fiber bundle and was delivered to the imaging target. The angle of incidence of the beam was 30 degrees relative to the imaging plane. The optical fluences were approximately 50 mJ/cm$^2$ for the 1,064 nm wavelength at 10 Hz PRF, 30 mJ/cm$^2$ at 20 Hz PRF, and 6 mJ/cm$^2$ at 100 Hz PRF, which were less than the safety limit set by the American National Standards Institute (ANSI) (100 mJ/cm$^2$, 50 mJ/cm$^2$, 10 mJ/cm$^2$, respectively, and 1000 mW/cm$^2$)[22].

To synchronize the system, the laser's external trigger was used to trigger the DAQ for photoacoustic signal acquisition. For each laser pulse, we acquired both the signal from the ultrasound probe surface and the signal from the imaging target. We then corrected the delay and jitter between the DAQ and the laser system using the acquired surface signal. The preprocessed raw signals were then backprojected to reconstruct the two-dimensional (2D) photoacoustic image.

**Data acquisition.** For the human imaging set-up, we prepared a portable water tank and placed it on top of the table. The ultrasound probe was mounted on a three-dimensional linear stage with its surface immersed in the water tank. The human hand/arm region was placed below the probe surface and coupled in water with the ultrasound probe. We implemented the photoacoustic imaging sequence to monitor the blood vessel in real-time. We achieved a frame rate of 10 Hz, 20 Hz or 100 Hz corresponding to the laser PRF. In the blood pressure cuff experiment, a blood pressure cuff device (GF Health Products, Inc) was placed on the upper arm region. The total acquisition took 30 s, with a baseline of ~5 s at the beginning, a cuff period of ~10 s following, and a release of ~15 s.

In the phantom validation experiments, the flow phantom was constructed using micro-renathane tubing (BrainTree Scientific). Each inlet channel had an inner diameter of 0.6 mm, and the outlet channel had an inner diameter of 1.0 mm. The phantom was perfused with 45% hematocrit whole bovine blood (QuadFive).

In the simulation study, we chose our red blood cell concentration to match a typical physiological value of ~ $5 \times 10^6$ red blood cells per microliter of blood in the total lumen region of the vessel (1 mm in diameter) for both the uniform blood and nonuniform blood cases. For the uniform blood case, the photoacoustic sources were randomly and uniformly distributed in the lumen region.

Conversely, in the case of nonuniform blood, the photoacoustic sources were randomly and nonuniformly distributed in the lumen region, with dense and sparse local regions. The simulation assumed a linear array ultrasound probe with the same bandwidth (13-24 MHz) and element arrangement (256 elements, 23 mm aperture size) as in the experimental case.

**Data processing.** To visualize the blood flow and obtain the vector flow map from the acquired raw photoacoustic signals, we used the following procedure for data processing as shown in Supplementary Fig. 7. We first applied a universal back projection image reconstruction algorithm[16] to the raw photoacoustic signals to reconstruct the images. We then logarithmically compressed the reconstructed structure images of the blood vessel (i.e., structure images in Fig. 1-5) to directly visualize the blood flow. For clearer blood flow visualization, we applied a singular value decomposition (SVD)-based spatiotemporal filter to the reconstructed images and then performed logarithmic compression on the filtered images. Lastly, we used a pixel-wise flow estimation method (see below) and noise floor filtering to obtain the velocity of the blood flow. For each pixel, the frame-to-frame velocity was estimated to form a 3D velocity structure with two spatial axes (ultrasound probe azimuthal direction $x$ and axial direction $z$) and one time axis (time $t$). The temporal standard deviation of each pixel's speed measurements was calculated to form a pixel-wise noise floor, below which frames with low speeds (i.e., measurements due to noise) were excluded. The final velocity map was constructed from averaging the remaining frames across the temporal domain.

In PAVT, a spatiotemporal filter is used to extract the blood component from the structural images. In our case, the spatiotemporal structure dataset has three dimensions with two spatial axes (ultrasound probe azimuthal direction $x$ and axial direction $z$) and one time axis (time $t$). We first reshape the 3D dataset into a 2D space-time matrix $S_{\text{structure}}(x, z, t)$. Then we use singular value decomposition (SVD)[34] to decompose the data matrix as follows:

$$S_{\text{structure}}(x, z, t) = \sum_{i=1}^{r} \sigma_i u_i(x, z) v_i^T(t), \qquad (1)$$

where $r$ is the rank of the data matrix, $\sigma_i$ is the $i^{th}$ singular value, $T$ is the conjugate transpose, $u_i(x, z)$ corresponds to the spatial domain, and $v_i(t)$ corresponds to the temporal domain. The static or slow-moving components (i.e., tissue) correspond to the first few larger singular values. Therefore, the relatively fast-moving blood components can be extracted as

$$S_{\text{blood}}(x,z,t) = S_{\text{structure}}(x,z,t) - \sum_{i=1}^{\lambda} \sigma_i u_i(x,z) v_i^T(t). \tag{2}$$

where $\lambda$ is the cutoff of the singular values for extracting the blood component. Finally, the filtered 2D space-time matrix $S_{\text{blood}}(x,z,t)$ is reshaped back to 3D with the same size as the original 3D dataset.

For flow estimation, the Farneback method[35] for optical flow was employed. Briefly, the Farneback method approximates pixel neighborhoods as polynomial expansions. By assuming a constant intensity of the displaced neighborhoods between adjacent frames, the displacement field is estimated as follows. Let the pixel neighborhood in Frame 1 at position vector $x$ be approximated as:

$$f_1(x) = x^T A_1 x + b_1^T x + c_1, \tag{3}$$

in which the coefficients $A_1$ (a symmetric matrix), $b_1$ (a vector), and $c_1$ (a scalar) are estimated by a weighted least squares fit of the signal. Now, let the displaced pixel neighborhood in Frame 2 be approximated as:

$$\begin{aligned} f_2(x) &= f_1(x-d) = (x-d)^T A_1 (x-d) + b_1^T(x-d) + c_1 \\ &= x^T A_1 x + (b_1 - 2A_1 d)^T x + d^T A_1 d - b_1^T d + c_1 \\ &= x^T A_2 x + b_2^T x + c_2, \end{aligned} \tag{4}$$

in which $d$ is the displacement vector to be estimated. By equating the coefficients

$$A_1 = A_2 \tag{5}$$

$$b_2 = b_1 - 2A_1 d \tag{6}$$

we solve for $d$:

$$d = -\frac{1}{2} A_1^{-1}(b_2 - b_1) \tag{7}$$

Implementations of the Farneback method can be found in OpenCV and the Computer Vision Toolbox of MATLAB.

In order to track the movement of the blood using PAVT, the imaging frame rate $f_{PRF}$ should satisfy the Nyquist sampling theorem[23,36], as determined by the equation below:

$$f_{\text{PRF}} > 2v/l, \tag{8}$$

where $l$ is the correlation length of the blood signals, and $v$ is the blood flow speed. In other words,

we must sample at least twice within the propagation time associated with the correlation length.

**Imaging protocols.** The experiments on human extremities were performed in a dedicated imaging room. All experiments were performed according to the relevant guidelines and regulations approved by the Institutional Review Board of the California Institute of Technology (Caltech). Eight healthy subjects were recruited from Caltech. Written informed consent was obtained from all the participants according to the study protocols.

**Reporting Summary.** Further information on research design is available in the Nature Research Reporting Summary linked to this article.

**References**

1. Won, R. Mapping blood flow. *Nat. Photonics* **5**, 393–393 (2011).

2. Rajan, V., Varghese, B., van Leeuwen, T. G. & Steenbergen, W. Review of methodological developments in laser Doppler flowmetry. *Lasers Med. Sci.* **24**, 269–283 (2009).

3. Leitgeb, R. A., Werkmeister, R. M., Blatter, C. & Schmetterer, L. Doppler Optical Coherence Tomography. *Prog. Retin. Eye Res.* **41**, 26–43 (2014).

4. Qureshi, M. M. *et al.* Quantitative blood flow estimation in vivo by optical speckle image velocimetry: publisher's note. *Optica* **8**, 1326–1326 (2021).

5. Boas, D. A. & Dunn, A. K. Laser speckle contrast imaging in biomedical optics. *J. Biomed. Opt.* **15**, 011109 (2010).

6. Cinotti, E. *et al.* Quantification of capillary blood cell flow using reflectance confocal microscopy. *Skin Res. Technol.* **20**, 373–378 (2014).

7. Wang, L. V. & Hu, S. Photoacoustic Tomography: In Vivo Imaging from Organelles to Organs. *Science* **335**, 1458–1462 (2012).

8. Wang, L. V. Tutorial on Photoacoustic Microscopy and Computed Tomography. *IEEE J. Sel. Top. Quantum Electron.* **14**, 171–179 (2008).

9. Yao, J. *et al.* High-speed label-free functional photoacoustic microscopy of mouse brain in action. *Nat. Methods* **12**, 407–410 (2015).

10. Kinnunen, M., Kauppila, A., Karmenyan, A. & Myllylä, R. Effect of the size and shape of a red blood cell on elastic light scattering properties at the single-cell level. *Biomed. Opt. Express* **2**, 1803–1814 (2011).

11. Wang, L. V. Multiscale photoacoustic microscopy and computed tomography. *Nat. Photonics* **3**, 503–509 (2009).

12. Brunker, J. & Beard, P. Velocity measurements in whole blood using acoustic resolution photoacoustic Doppler. *Biomed. Opt. Express* **7**, 2789–2806 (2016).

13. Yao, J. & Wang, L. V. Photoacoustic brain imaging: from microscopic to macroscopic scales. *Neurophotonics* **1**, 011003 (2014).

14. Guo, Z., Li, L. & Wang, L. V. On the speckle-free nature of photoacoustic tomography. *Med. Phys.* **36**, 4084–4088 (2009).

15. Pakdaman Zangabad, R. *et al.* Photoacoustic flow velocity imaging based on complex field decorrelation. *Photoacoustics* **22**, 100256 (2021).

16. Xu, M. & Wang, L. V. Universal back-projection algorithm for photoacoustic computed tomography. *Phys. Rev. E* **71**, 016706 (2005).

17. Lurie, F., Kistner, R. L., Eklof, B. & Kessler, D. Mechanism of venous valve closure and role of the valve in circulation: a new concept. *J. Vasc. Surg.* **38**, 955–961 (2003).

18. Szabo, T. L. *Diagnostic ultrasound imaging: inside out*. (Academic press, 2004).

19. Tanter, M. & Fink, M. Ultrafast imaging in biomedical ultrasound. *IEEE Trans. Ultrason. Ferroelectr. Freq. Control* **61**, 102–119 (2014).

20. Errico, C. *et al.* Ultrafast ultrasound localization microscopy for deep super-resolution vascular imaging. *Nature* **527**, 499–502 (2015).

21. Wang, L. V. & Wu, H. *Biomedical optics: principles and imaging*. (John Wiley & Sons, 2012).

22. ANSI Z136.1-2014 - American National Standard for Safe Use of Lasers. https://webstore.ansi.org/Standards/LIA/ANSIZ1362014.

23. Grenander, U. The Nyquist frequency is that frequency whose period is two sampling intervals. *Probab. Stat. Harald Cramér Vol.* **434**, (1959).

24. Lee, Y., Kang, J. & Yoo, Y. Automatic dynamic range adjustment for ultrasound B-mode imaging. *Ultrasonics* **56**, 435–443 (2015).

25. Fernández-Colino, A. & Jockenhoevel, S. Advances in engineering venous valves: the pursuit of a definite solution for chronic venous disease. *Tissue Eng. Part B Rev.* **27**, 253–265 (2021).

26. Petrila, T. & Trif, D. *Basics of fluid mechanics and introduction to computational fluid dynamics*. vol. 3 (Springer Science & Business Media, 2004).

27. Murray, C. D. The Physiological Principle of Minimum Work. *Proc. Natl. Acad. Sci. U. S. A.* **12**, 207–214 (1926).

28. Li, L. *et al.* Single-impulse panoramic photoacoustic computed tomography of small-animal whole-body dynamics at high spatiotemporal resolution. *Nat. Biomed. Eng.* **1**, 1–11 (2017).

29. Keys, A. The oxygen saturation of the venous blood in normal human subjects. *Am. J. Physiol.-Leg. Content* **124**, 13–21 (1938).

30. Na, S., Zhang, Y. & Wang, L. V. Cross-Ray Ultrasound Tomography and Photoacoustic Tomography of Cerebral Hemodynamics in Rodents. *Adv. Sci.* 2201104 (2022).

31. Yao, J., Maslov, K. I. & Wang, L. V. In vivo photoacoustic tomography of total blood flow and potential imaging of cancer angiogenesis and hypermetabolism. *Technol. Cancer Res. Treat.* **11**, 301–307 (2012).

32. Na, S. *et al.* Massively parallel functional photoacoustic computed tomography of the human brain. *Nat. Biomed. Eng.* 1–9 (2021) doi:10.1038/s41551-021-00735-8.

33. Beebe-Dimmer, J. L., Pfeifer, J. R., Engle, J. S. & Schottenfeld, D. The Epidemiology of Chronic Venous Insufficiency and Varicose Veins. *Ann. Epidemiol.* **15**, 175–184 (2005).

34. Demené, C. *et al.* Spatiotemporal Clutter Filtering of Ultrafast Ultrasound Data Highly Increases Doppler and fUltrasound Sensitivity. *IEEE Trans. Med. Imaging* **34**, 2271–2285 (2015).

35. Farnebäck, G. Two-frame motion estimation based on polynomial expansion. in *Scandinavian conference on Image analysis* 363–370 (Springer, 2003).

36. Leis, J. W. *Digital signal processing using MATLAB for students and researchers*. (John Wiley & Sons, 2011).

**Data availability**

The data that support the findings of this study are provided within the paper and its supplementary material.

**Code availability**

The reconstruction algorithm and data processing methods can be found in the Methods. The reconstruction code is not publicly available because it is proprietary and is used in licensed technologies.

**Acknowledgments**

We thank Konstantin Maslov, Lei Li, and Rui Cao for their discussions about the flow mechanism. We thank Steven L. Spitalnik and Paul Buehler for their discussions about blood physiology. We thank Sam Davis and Byullee Park for discussion on the potential improvement of flow visualization. This work was sponsored by the United States National Institutes of Health (NIH) grants U01 EB029823 (BRAIN Initiative) and R35 CA220436 (Outstanding Investigator Award).

**Contributions**

L.V.W., Y.Z., and J.O.G. designed the study. Y.Z. and J.O.G. built the system, analyzed the data and wrote the manuscript with input from all of the authors. Y.Z., J.O.G. and A.K. performed the experiments. L.V.W., Y.Z., J.O.G., and A.K. interpreted the data. L.V.W. supervised the study and revised the manuscript.

**Competing interests**

L.V.W. has a financial interest in Microphotoacoustics Inc., CalPACT LLC, and Union Photoacoustic Technologies Ltd., which, however, did not support this work.

**Supplementary figures**

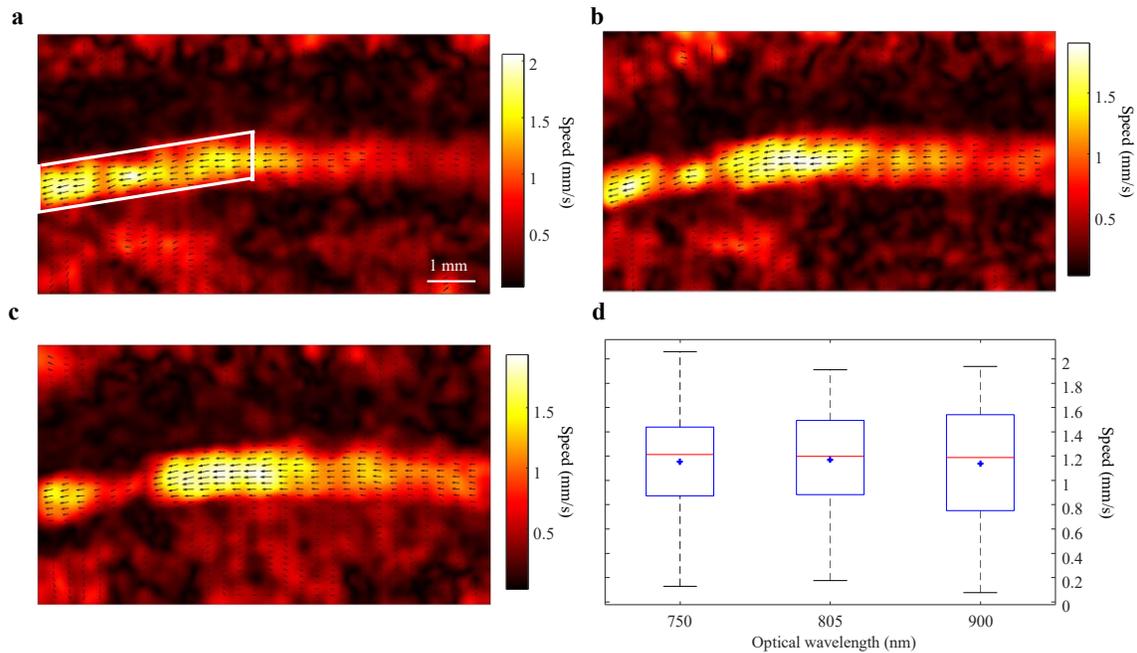

**Supplementary Fig. 1 | PAVT at multiple wavelengths.** A blood vessel in the palmar region was imaged at three different optical wavelengths in the near-infrared spectral region. **a-c,** Vector flow maps for the vessel imaged at 750, 805, and 900 nm, respectively. **d,** Box-and-whisker plot of measured speeds at each wavelength. The mean ± standard deviation (in mm/s) speeds for wavelengths of 705, 805, and 900 nm were 1.15 ± 0.39, 1.17 ± 0.40, and 1.14 ± 0.47, respectively ($n$ = 9971). The red horizontal line shows the median; the boxes indicate the 25$^{th}$ and 75$^{th}$ percentiles; the blue cross shows the mean value. The White box indicates the region for speed calculations.

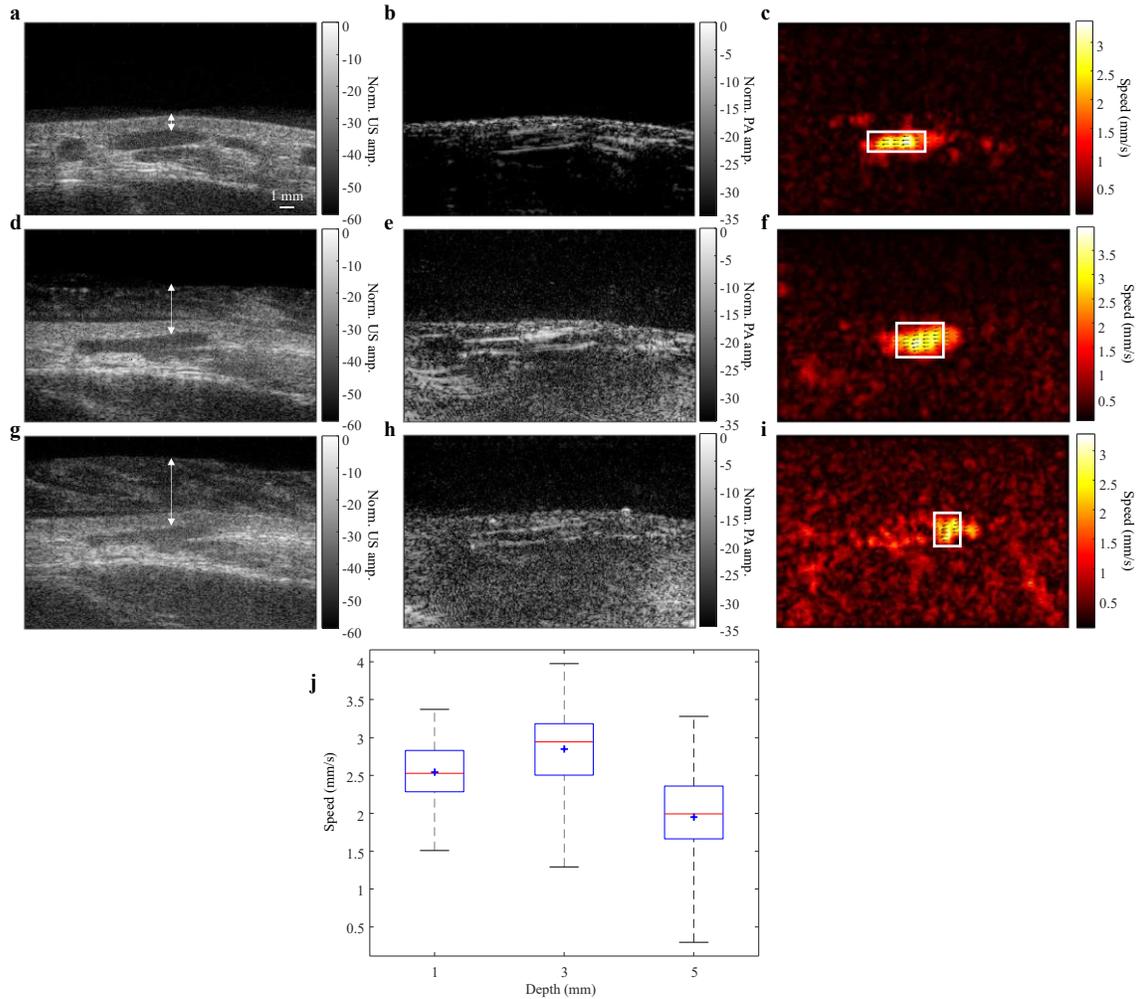

**Supplementary Fig. 2 | Measuring blood flow in the skin with overlying chicken tissue to demonstrate penetration.** A light-scattering slab of chicken tissue with varying thicknesses was placed on top of a metacarpal vessel (distance from the skin surface to the top of the vessel ~1 mm). The ultrasound image, photoacoustic image, and PAVT vector flow maps for the vessel **a-c**, with no chicken tissue (effective depth ~1 mm), **d-f**, with 2 mm of chicken tissue placed on top of the skin (effective depth ~3 mm), **g-i**, with 4 mm of chicken tissue placed on top of the skin (effective depth ~5 mm). **j,** Box-and-whisker plot of measured speeds at each depth. The mean ± standard deviation (in mm/s) speeds and sample number for depths of 1, 3 and 5 were 2.54 ± 0.39 ($n$ = 997), 2.85 ± 0.49 ($n$ = 1629, and 1.95 ± 0.61 ($n$ = 930), respectively. The red horizontal line shows the median; the boxes indicate the 25$^{th}$ and 75$^{th}$ percentiles; the blue cross shows the mean value. White boxes indicate the regions for speed calculations.

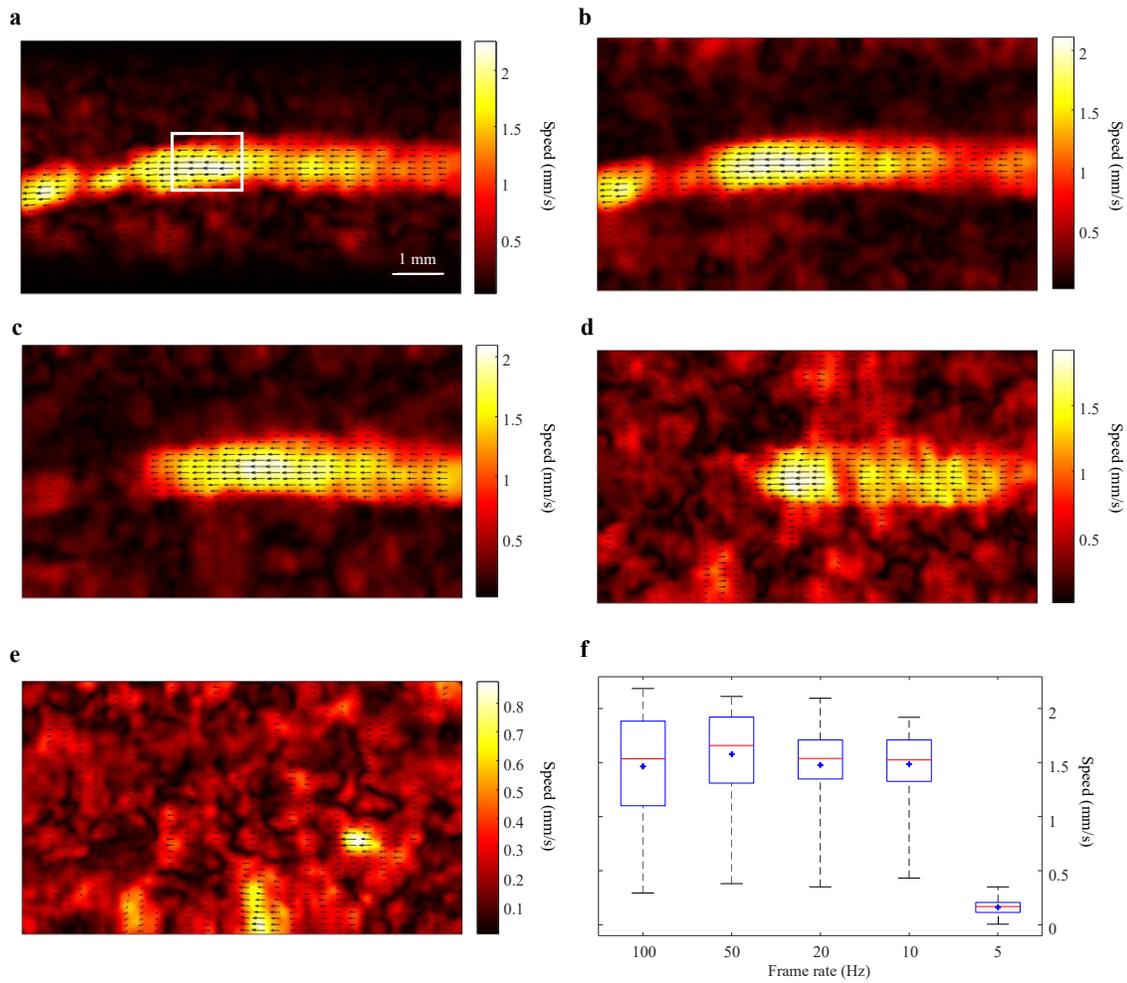

**Supplementary Fig. 3 | Effect of downsampling.** Images of a palmar vessel were acquired for one second at a pulse repetition frequency (PRF) of 100 Hz. **a-e,** Vector flow maps for sampling the images at frame rates of 100, 50, 20, 10, and 5 Hz, respectively. **f,** Box-and-whisker plot of measured speeds at each frame rate. The mean ± standard deviation (in mm/s) speeds for frame rates of 100, 50, 20, 10, and 5 Hz were 1.47 ± 0.49, 1.58 ± 0.40, 1.48 ± 0.33, 1.49 ± 0.28, and 0.16 ± 0.07, respectively ($n$ = 2952). The red horizontal line shows the median; the boxes indicate the 25$^{th}$ and 75$^{th}$ percentiles; the blue cross shows the mean value. White box indicates the region for speed calculations.

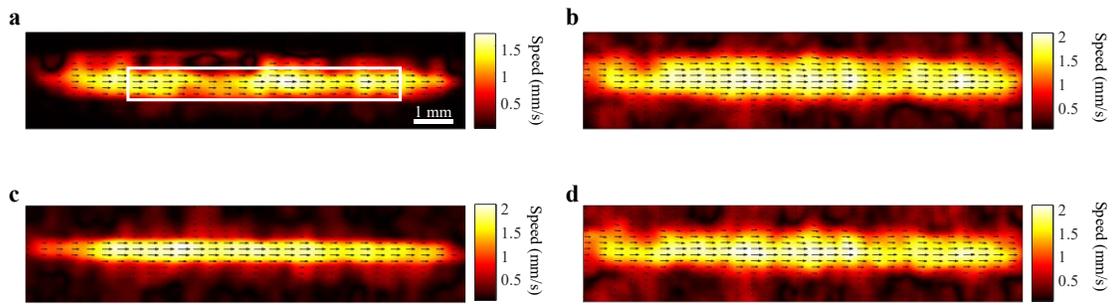

**Supplementary Fig. 4 | Effect of SVD and logarithmic compression.** Images of a blood phantom with a mean flow speed of 1.645 mm/s were acquired as a ground truth. **a-b,** Vector flow maps obtained from processing the images on a linear scale with and without SVD processing, respectively. **c-d** Vector flow maps obtained from processing the images on a logarithmic scale with and without SVD processing, respectively. The mean (in mm/s) speeds across the region indicated by the white box for **a-d** were 1.154, 1.758, 1.622, and 1.640 with relative errors 29, 6.9, 1.4, and 0.3 % respectively (*n* = 5599).

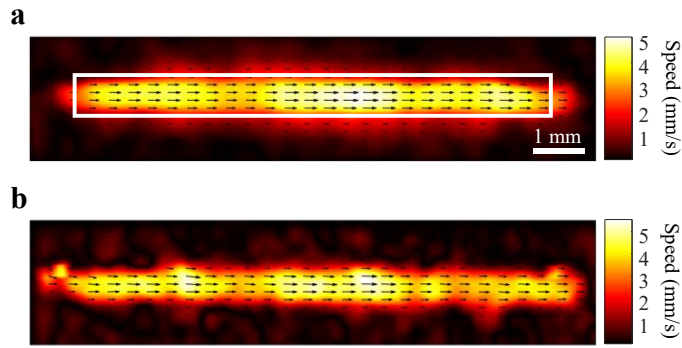

**Supplementary Fig. 5 | Effect of noise floor filtering.** Images of a blood phantom with a mean flow speed of 4.11 mm/s were acquired as a ground truth. **a,** Vector flow map obtained from averaging across all frames for each pixel. **b,** Vector flow map obtained from averaging after performing noise floor filtering for each pixel. The mean (in mm/s) and sample number across the region indicated by the white box for **a-b** were 3.93 ($n = 6232$) and 4.12 ($n = 6232$) with relative errors of 4.5 and 0.9 %, respectively.

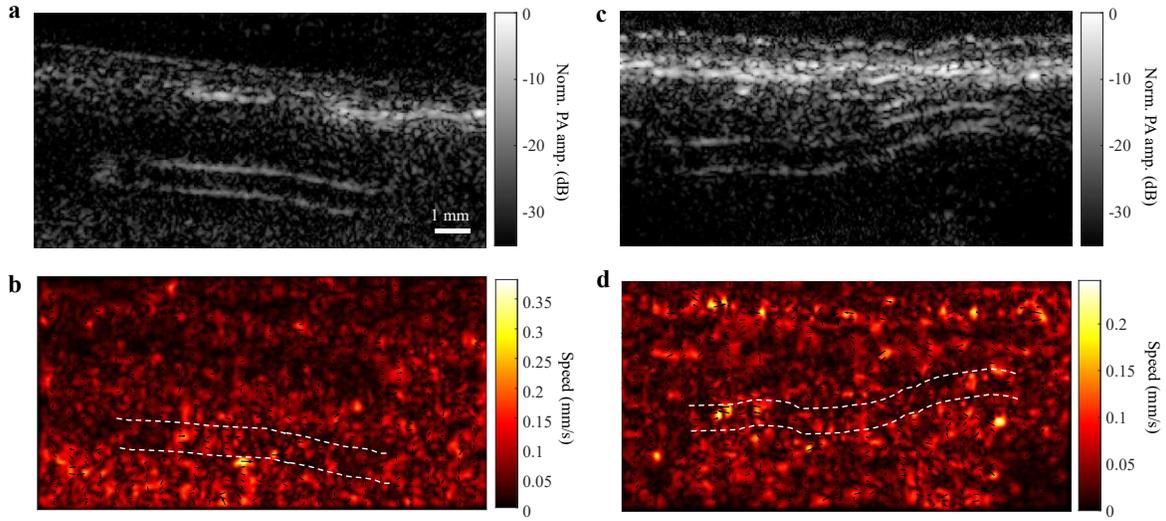

**Supplementary Fig. 6 | Measuring blood flow in arteries.** Two blood vessels in the hand region were imaged. **a-b**, Structure and vector flow maps for an artery at a depth of 3 mm. **c-d**, structure and vector flow maps for another artery at a depth of 2.5 – 3 mm. White dashed lines indicate the vessel region.

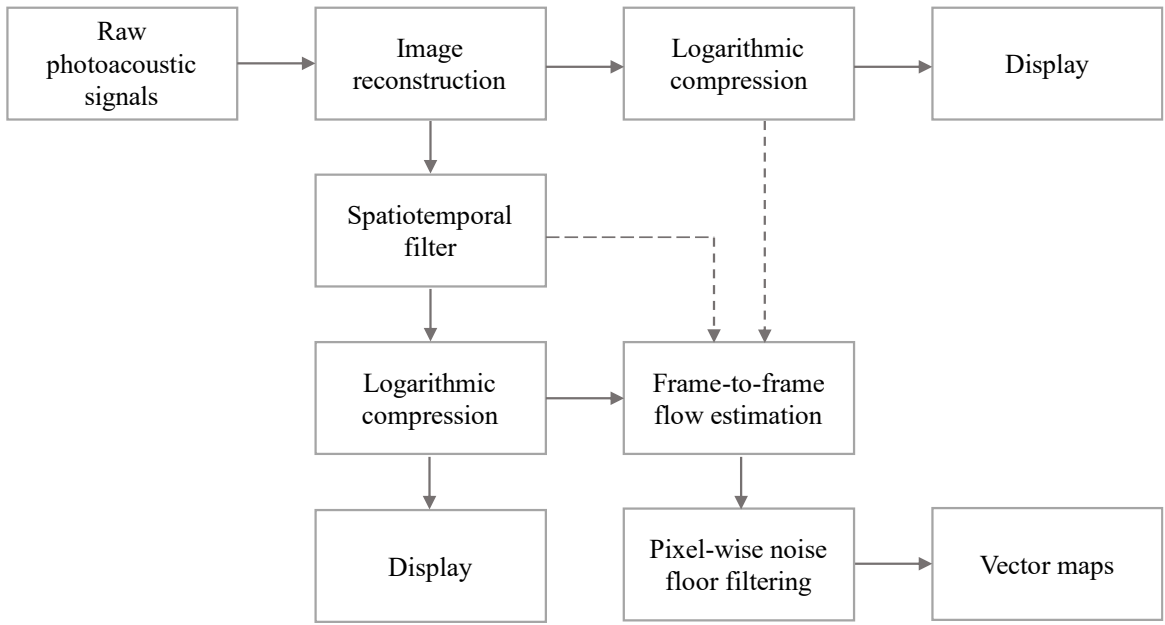

**Supplementary Fig. 7 | Data processing flowchart.** We first applied an image reconstruction algorithm to the raw photoacoustic signals to obtain the reconstructed images. We then performed logarithmic compression on the reconstructed images for obtaining the structure images of the blood vessel and to directly visualize the blood flow. Alternatively, for clearer blood flow visualization, we applied a singular value decomposition (SVD)-based spatiotemporal filter to the reconstructed images and then performed logarithmic compression on the filtered images. Lastly, we used a frame-to-frame flow estimation algorithm and pixel-wise noise floor filtering to obtain the vector maps of the blood flow.